\documentclass{elsart}
\usepackage{amsfonts}
\usepackage{amssymb}
\usepackage{amsmath}
\usepackage[all]{xy}
\begin{document}

\begin{frontmatter}

\title{From screening to confinement in a Higgs-like model}

\author{Patricio Gaete\thanksref{cile}}
\thanks[cile]{e-mail address: patricio.gaete@usm.cl}
\address{Departamento de F\'{\i}sica, Universidad T\'ecnica
Federico Santa Mar\'{\i}a, Valpara\'{\i}so, Chile}

\author{Euro Spallucci\thanksref{infn}}
\thanks[infn]{e-mail address: spallucci@ts.infn.it }
\address{Dipartimento di Fisica Teorica, Universit\`a di Trieste
and INFN, Sezione di Trieste, Italy}

\begin{abstract}
We investigate a recently proposed Higgs-like model (arXiv:0811.4423 [hep-th]),
in the framework of a gauge-invariant but path-dependent variables formalism.
We compute the static potential between test charges in a condensate of
scalars and fermions.\\
In the case of charged massive scalar we recover the screening potential.
On the other hand, in the Higgs case,  with a  ''tachyonic" mass term and a
quartic potential in the Lagrangian, unexpected features are found. It is
observed that the interaction energy is the sum of an effective-Yukawa  and
a linear potential, leading to the confinement of static charges.
\end{abstract}
\end{frontmatter}

\section{Introduction}
It is a well known fact that, despite an intense effort for more than three
decades, a full understanding of the QCD vacuum structure and color
confinement mechanism are still lacking.  String theory $AdS/CFT$ duality
\cite{Maldacena:2002eg} is the most promising technical framework to face
non-perturbative $QCD$ effects. Despite an intensive activity in this advanced
research sector, a final word is still to come. In the meanwhile, phenomenological
models still offer a proper platform for understanding different features
of the physics of confinement. In this context it may be useful to  recall that
the phenomenon of condensation has provided valuable insight and theoretical
guidance into this problem. For instance, in the illustrative  picture of dual
superconductivity \cite{Nambu:1974zg,Mandelstam:1976tq,'tHooft:1975pr}, where
it is conjectured that the QCD vacuum behaves as a dual-type II superconductor.
More explicitly, because of the condensation of magnetic monopoles, the
chromoelectric field acting between $q\overline q$ pair is squeezed into
strings, and the nonvanishing string tension represents the proportionality
constant in the linear potential. In passing let us also mention a theory of
antisymmetric tensor fields that results from the condensation of topological
defects \cite{Seo:1981tm,Mansfield:1985bp,Guendelman:1989dp,Quevedo:1996tx,Ansoldi:2000qs} as a
consequence of the Julia-Toulouse mechanism, which also leads to confinement
of static charges \cite{Gaete:2004dn}.
\\
Condensation of charged scalars and its physical
consequences such as Friedel oscillations and the strong screening of
an electric charge have created a lot of interest in recent times
\cite{Gabadadze:2007si,Gabadadze:2008sq,Gabadadze:2008mx,Gabadadze:2008pj}.
Much of this work has been motivated by the possibility of describing condensed
helium-4 nuclei in an electron background in white dwarf cores. In particular,
a Lorentz-violating Higgs-like effective Lagrangian has been considered.
The crucial ingredient of this development is to introduce a nonzero vacuum
expectation value for the fermion field which permits to
realize the condensation of the helium-4. As a result, it was shown a
strong screening between probe charges placed in the condensate.
It should be noted that the condensate characterizes the new vacuum of the
theory with striking consequences over the different phases of the pure gauge
sector of the proposed model. Given the relevance of these studies, it is of
interest to enrich our understanding of the physical consequences presented
by condensates. Thus, our purpose here is to further explore the impact of
condensates on physical observables. Of special interest will be to provide an
additional independent demonstration on the screening in the condensate,
based in a Hamiltonian framework.\\

As we mentioned above, our concern in this work is to provide a different
calculation of the interaction energy between pointlike sources in the
framework of the recently proposed Lorentz-violating Higgs-like effective model
\cite{Gabadadze:2008pj,Dolgov:2008pe}. To this end we will use the gauge-invariant but 
path-dependent
variables formalism \cite{Gaete:1999iy}, which is a physically-based alternative
to the Wilson loop approach. When we compute in this way the static potential for
the model described in \cite{Gabadadze:2008pj}, in the case of a "right-sign" mass
term $m_H^2 \phi ^ *  \phi$, we obtain an effective - Yukawa potential,  which in the
$m_\gamma   \ll M$ approximation is identical to the one encountered in
\cite{Gabadadze:2008pj}. On the other hand, in the case of a "wrong-sign''
mass term $- m_H^2 \phi ^ *  \phi$, the result of this calculation is new and rather
unexpected. We show that the interaction energy is the sum of an effective -Yukawa and
a \textit{linear potential}, leading to the confinement of static charges. It is
interesting to note that the above static profile is analogous to that encountered
in both Abelian and non-Abelian models. Remarkable examples are:  theory of
antisymmetric tensor fields that results from the condensation of topological defects
as a consequence of the Julia-Toulouse mechanism \cite{Gaete:2004dn}; gauge theory
with a pseudoscalar coupling, in the case that there is a constant magnetic
strength expectation value \cite{Gaete:2004ga}; gauge theory which
includes the mixing between the familiar photon $U(1)_{QED}$ and a second
massive gauge field living in the so-called hidden-sector $U(1)_h$
\cite{Gaete:2008qx}. On the non-Abelian side, we find the case of gluodynamics in
curved space-time \cite{Gaete:2007zn}, and of non-Abelian gauge theory
with a mixture of pseudoscalar and scalar couplings, in the case where a
constant chromoelectric, or chromomagnetic, strength expectation value is
present \cite{Gaete:2007fu}. These connections thus provide us
with a new kind of "duality" among diverse models and allow us to exploit
this equivalence in concrete calculations, as we are going to illustrate.

\section{Scalar condensation}

We turn now to the problem of obtaining the interaction energy
between static point-like sources for a Lorentz-violating Higgs-like
effective model. To do this, we shall compute the expectation value
of the energy operator $H$ in the physical state $|\Phi\rangle$ describing
the sources, which we will denote by ${\langle H\rangle}_\Phi$.  However,
before going to the derivation of the interaction potential, we will describe
the salient features of this model in the familiar language of standard
quantum field theory. For this purpose we restrict our attention to scalar
electrodynamics in the presence of a fermion background density:

\begin{equation}
\mathcal{ L} =  - \frac{1}{4}F_{\mu \nu }^2  + \left| {D_\mu  \phi }
\right|^2  - m_H^2 \phi ^ *  \phi + \overline \psi  \left( {i\gamma ^\mu  D_\mu
- M_f } \right)\psi,   \label{cond05}
\end{equation}

$\phi$ is a charged, massive, scalar field; $A_\mu$ is a $U\left(\,1\,\right)$
gauge potential and  $\psi$ is an ``heavy'' fermion. The covariant derivative is
defined as usual
\begin{equation}
D_\mu \equiv \partial_\mu -i e A_\mu.
\end{equation}
Let us remark that $m_H^2>0$ is a ``right sign'' mass term  and we have
not included any self-interaction for the scalar field. Indeed, the condensation
mechanism we are going to consider is quite different from spontaneous symmetry
breaking and relies on the presence of a non-trivial fermion density in the ground
state. In order to implement this setting we suppose fermions to be so heavy that
they cannot be excited in the low energy regime we are considering. Thus, the Dirac
kinetic term can be neglected and the whole fermion sector of the model reduces
to a constant background density  $J^0$ coupled to $A_\mu$
\begin{equation}
 \overline \psi \gamma ^\mu \psi\longrightarrow \delta^\mu_0\, J^0.
\end{equation}
In such a case,
\begin{equation}
\mathcal{ L}\to - \frac{1}{4}F_{\mu \nu }^2  + \left| {D_\mu  \phi } \right|^2
- m_H^2 \phi ^ *  \phi +e J^0  \delta^\mu_0\, A_\mu. \label{cond05a}
\end{equation}

In Eq.(\ref{cond05a}) $J^\mu$ acts as an ``external source'' for the time component
of the gauge potential. From this point of view, the fermionic condensate breaks
Lorentz invariance by giving a constant vacuum expectation value to $A_0$  as we are
going to show in a while. \\
The field equations obtained by varying (\ref{cond05a}) with respect to $A_\mu$ and
$\phi^\ast$ read

\begin{eqnarray}
&& \partial_\mu F^{\mu\nu} +2e^2 A^\nu |\phi|^2 = eJ^\nu_s + e J^0
\delta^\nu_0\label{f1}\\
&& \left(\, D_\mu D^\mu + m^2_H \,\right)\phi =0\label{f2}
\end{eqnarray}

where $ J^\nu_s\equiv i\phi^\ast\partial^\nu \phi -i\phi\partial^\nu \phi^\ast $.\\
The presence of a homogeneous source $J^0$ allows the existence of constant classical
solutions of the form

\begin{eqnarray}
 && \phi=\phi^\ast=\mathrm{const.}\equiv \phi_0\ne 0\ , \label{cs1}\\
 && A_\mu \equiv \frac{\mu_s}{e}\,\delta^0_\mu.  \label{cs2}
\end{eqnarray}

The two integration constants $\phi_0$, $\mu_s$ are determined by inserting (\ref{cs1}),
(\ref{cs2}) in (\ref{f1}), (\ref{f2}):
\begin{eqnarray}
 && 2e^2 A^\nu \phi^2_0 =   e J^0  \delta^\nu_0,  \label{f11}\\
 && \left(\, e^2\, A^2 - m^2_H \,\right)\phi_0 =0. \label{f22}
\end{eqnarray}

Eq.(\ref{f1}) and (\ref{cs2}) determine $\mu_s$

\begin{equation}
A^2 = \frac{m^2_H}{e^2}=\frac{\mu_s^2}{e^2}\longrightarrow \mu_s=m_H
\end{equation}

Eq.(\ref{f2}) and (\ref{cs1}) fix $\phi_0$

\begin{equation}
A_\mu =  \frac{J^0}{2e\phi_0^2}\,  \delta_{\mu\,0}=\frac{m_H}{e} \,  \delta_{\mu\,0}
\longrightarrow \phi_0=\sqrt{\frac{J^0}{2m_H}}
\end{equation}

In summary, the  ground state of the system is described by the classical solution:

\begin{eqnarray}
 && \overline{\psi}_0 \gamma ^\mu \psi_0= \delta^\mu_0\, J^0, \label{g1}\\
 && \phi_0=\sqrt{\frac{J^0}{2m_H}}, \label{g2}\\
 && A_{(0)\,\mu }=\frac{m_H}{e} \,  \delta_\mu ^{\,0}. \label{g3}
\end{eqnarray}

While fermions are ``frozen'' into the ground state, both $\phi$ and $A_\mu$ are
subject to quantum fluctuations. More exactly, the modulus of $\phi$ oscillates and
its phase has been gauged to zero. By splitting fields into a background value and a
dynamical part:

\begin{eqnarray}
 && \phi =\phi^*= \phi_0 +\frac{1}{\sqrt{2}}\eta\left(\, x\,\right), \label{fluc1}\\
&& A_\mu =\frac{m_H}{e} \,  \delta^0_\mu + b_\mu\left(\, x\,\right), \label{fluc2}
\end{eqnarray}

we expand the Lagrangian up to quadratic terms in the fluctuations

\begin{equation}
\mathcal{ L}^{(2)} =  - \frac{1}{4}f_{\mu \nu }^2  + \frac{1}{2}\left( {\partial _\mu
\eta } \right)^2
+ \frac{1}{2}m_\gamma ^2 b_\mu ^2  + 2m_H m_\gamma b_0\,  \eta\ . \label{cond05b}
\end{equation}
Here $f_{\mu \nu } \equiv \partial _\mu  b_\nu   - \partial _\nu  b_\mu$ and
$m^2_\gamma\equiv 2e^2\phi_0^2$.\\
It is important to remark that the scalar fluctuation $\eta$ is massless due to a
cancelation between the original mass term $m^2_H$ and a quadratic term from the gauge
covariant kinetic term. In the case of the Higgs field the same two terms sum up leading
to massive fluctuations and a different static potential. We are going to discuss this
case in the last part of the paper.\\
Next, integrating out the $\eta$ field induces an effective theory for the  $b_\mu$ field,
that is,
\begin{equation}
\mathcal{ L} =  - \frac{1}{4}f_{\mu \nu }^2  + \frac{1}{2}m_\gamma ^2 b_\mu ^2
+ 2m_H^2 m_\gamma^2\, b_0\, \frac{1}{\Delta}\,b_0, \label{cond10}
\end{equation}\\
which is a Maxwell-Proca-like theory with a manifestly Lorentz violating term.
It can be worth to remark that the mixing  of such a term with the Lorentz invariant
ones is the key element to reproduce the correct screening potential.\\
In order to restore the gauge invariance in (\ref{cond10}), we shall use the Hamiltonian
formalism for constrained systems because it leads directly to a physical Hamiltonian,
as we are going to illustrate in the next section.

\section{Interaction energy}
As already mentioned, our immediate objective is to calculate the interaction energy
between external probe sources for the model under consideration. To do this, our
first undertaking is to restore the gauge invariance in (\ref{cond10}). For this we
shall use the Hamiltonian formalism for constrained systems along the lines of
Ref. \cite{GaeteGuenSpa07}. To achieve this end we first note that the Lagrangian
(\ref{cond10}) may be rewritten as
\begin{equation}
\mathcal{ L} =  - \frac{1}{4}f_{\mu \nu }^2  + \frac{1}{2}b_\mu  m^2 b^\mu
- \frac{1}{2}b_i \frac{{\left( {2m_H m_\gamma  } \right)^2 }}{\Delta }b^i , \label{cond15}
\end{equation}
where $m^2  \equiv m_\gamma ^2 \left( {1 + \frac{{4m_H^2 }}{\Delta }} \right)$.
The canonically conjugate are $\Pi ^0=0$ and $\Pi ^i   =  - f^{0i}$. This leads us
to the canonical Hamiltonian,
\begin{equation}
H = \int {d^3 x} \left\{ { - b_0 \left( {\partial _i \Pi ^i  + \frac{{m^2 }}{2}b^0 }
\right) - \frac{1}{2}\Pi _i \Pi ^i  + \frac{1}{4}f_{ij} f^{ij}  - \frac{1}{2}b_i
\left( {m^2  - \frac{{\left( {2m_H m_\gamma  } \right)}}{\Delta }} \right)b^i }
\right\}, \label{cond20}
\end{equation}
Requiring the primary constraint  $\Pi ^0=0$ to be preserved in time yields the
following secondary constraint
\begin{equation}
\Gamma \left( x \right) \equiv \partial _i \Pi ^i  + m^2 b^0  = 0. \label{cond25}
\end{equation}
The above result reveals the second class nature of the constraints, as expected for
a theory with an explicit mass term which breaks the gauge invariance. Next, as was
explained in Ref. \cite{GaeteGuenSpa07}, we enlarge the original phase space by
introducing a canonical pair of fields $\theta$ and $ \Pi _\theta $. Accordingly,
a new set of constraints can be defined in this extended space:
\begin{equation}
\Lambda _1  \equiv \Pi _0  + m^2 \theta, \label{cond30a}
\end{equation}
and
\begin{equation}
\Lambda _2  \equiv \Gamma  + \Pi _\theta. \label{cond30b}
\end{equation}
It is simple to see that the new constraints are first class and in this way restore
the gauge symmetry of the theory under consideration. Therefore the new effective
Lagrangian, after integrating out the $\theta$ field, reads
\begin{equation}
\mathcal{ L} =  - \frac{1}{4}f_{\mu \nu } \left[ {1 + \frac{{m_\gamma ^2 }}
{\Delta }\left( {1 + \frac{{4m_H^2 }}{\Delta }} \right)} \right]f^{\mu \nu }.
\label{cond35}
\end{equation}
We observe that to get this new theory we have ignored the last term in (\ref{cond15})
because it add nothing to the static potential calculation, as we will show it below.
In other words, this new effective theory provide us with a suitable starting point
to study the interaction energy.

Having characterized the theory under study, we can now compute the interaction
energy. To obtain the corresponding Hamiltonian, we must carry out the quantization
of this theory. The Hamiltonian analysis starts with the computation of the canonical
momenta $\Pi ^\mu   =  - \left[ {1 + \frac{{m_\gamma ^2 }}{\Delta }\left( {1 +
\frac{{4m_H^2 }}{\Delta }} \right)} \right]f^{0\mu }$, which produces the usual primary
constraint $\Pi ^0  = 0$ and $ \Pi ^i  =  - \left[ {1 + \frac{{m_\gamma ^2 }}{\Delta }
\left( {1 + \frac{{4m_H^2 }}{\Delta }} \right)} \right]f^{0i}$. The canonical Hamiltonian
is then
\begin{equation}
H_C  = \int {d^3 x} \left\{ { - b_0 \partial _i \Pi ^i  - \frac{1}{2}\Pi _i
\left[ {1 + \frac{{m_\gamma ^2 }}{\Delta }\left( {1 + \frac{{4m_H^2 }}{\Delta }}
\right)} \right]^{ - 1} \Pi ^i  + \frac{1}{4}f_{ij} f^{ij} } \right\}. \label{cond40}
\end{equation}
Time conservation of the primary constraint yields the usual Gauss
constraint $\Gamma_1 \left( x \right) \equiv \partial _i \Pi ^i=0$.
Note that the time stability of this constraint does not induce
further constraints. Consequently, the extended Hamiltonian that
generates translations in time then reads $H = H_C + \int {d^3
}x\left( {c_0 \left( x \right)\Pi _0 \left( x \right) + c_1 \left(
x\right)\Gamma _1 \left( x \right)} \right)$. Here $c_0 \left(
x\right)$ and $c_1 \left( x \right)$ are arbitrary Lagrange
multipliers. It should be noted that $\dot{b}_0 \left( x \right)=
\left[ {b_0 \left( x \right),H} \right] = c_0 \left( x \right)$,
which is an arbitrary function. Since $ \Pi^0 = 0$ always, neither $
b^0 $ nor $ \Pi^0 $ are of interest in describing the system and may
be discarded from the theory. Thus the Hamiltonian is now given as
\begin{equation}
H = \int {d^3 x} \left\{ {  c(x) \partial _i \Pi ^i  - \frac{1}{2}\Pi _i
\left[ {1 + \frac{{m_\gamma ^2 }}{\Delta }\left( {1 + \frac{{4m_H^2 }}
{\Delta }} \right)} \right]^{ - 1} \Pi ^i  + \frac{1}{4}f_{ij} f^{ij} }
\right\}, \label{cond45}
\end{equation}
where $c(x) = c_1 (x) - b_0 (x)$.

In order to break the gauge freedom of the theory, it is necessary to
impose one gauge constraint such that the full set of constraints become
second class. A particularly convenient choice is found to be
\begin{equation}
\Gamma _2 \left( x \right) \equiv \int\limits_{C_{\xi x} } {dz^\nu }
b_\nu \left( z \right) \equiv \int\limits_0^1 {d\lambda x^i } b_i
\left( {\lambda x} \right) = 0,     \label{cond50}
\end{equation}
where  $\lambda$ $(0\leq \lambda\leq1)$ is the parameter describing
the spacelike straight path $ x^i = \xi ^i  + \lambda \left( {x -
\xi } \right)^i $, and $ \xi $ is a fixed point (reference point).
There is no essential loss of generality if we restrict our
considerations to $ \xi ^i=0 $. The choice (\ref{cond50}) leads to
the Poincar\'e gauge \cite{Gaete:1999iy}. As a consequence, we can
now write down the only nonvanishing Dirac bracket for the canonical
variables
\begin{equation}
\left\{ {b_i \left( x \right),\Pi ^j \left( y \right)} \right\}^ *
=\delta{ _i^j} \delta ^{\left( 3 \right)} \left( {x - y} \right) -
\partial _i^x \int\limits_0^1 {d\lambda x^j } \delta ^{\left( 3
\right)} \left( {\lambda x - y} \right). \label{cond55}
\end{equation}

We have finally assembled the tools to determine the interaction
energy for the model under consideration. As mentioned before, in
order to accomplish this purpose we will calculate the expectation
value of the energy operator $H$ in the physical state
$|\Phi\rangle$. Now we recall that the physical state $|\Phi\rangle$
can be written as
\begin{equation}
\left| \Phi  \right\rangle  \equiv \left| {\overline \Psi  \left(
\bf y \right)\Psi \left( {\bf y}\prime \right)} \right\rangle
= \overline \psi \left( \bf y \right)\exp \left(
{iq\int\limits_{{\bf y}\prime}^{\bf y} {dz^i } b_i \left( z \right)}
\right)\psi \left({\bf y}\prime \right)\left| 0 \right\rangle,
\label{cond60}
\end{equation}
where the line integral is along a spacelike path on a fixed time
slice, and $\left| 0 \right\rangle$ is the physical vacuum state.

Now, using the previous Hamiltonian structure, and since the
fermions are taken to be infinitely massive (static) we can
substitute $\Delta$ by $-\nabla^{2}$ in Eq. (\ref{cond45}). In such a
case $\left\langle H \right\rangle _\Phi$ reduces to
\begin{equation}
\left\langle H \right\rangle _\Phi   = \left\langle H \right\rangle
_0 + \left\langle H \right\rangle _\Phi ^{\left( 1 \right)},
\label{cond65}
\end{equation}
where $\left\langle H \right\rangle _0  = \left\langle 0
\right|H\left| 0 \right\rangle$, and the $\left\langle H \right\rangle
 _\Phi ^{\left( 1 \right)}$ term is given by
\begin{equation}
\left\langle H \right\rangle _\Phi ^{\left( 1 \right)}  = \left\langle \Phi
\right|\int {d^3 x} \left\{ { - \frac{1}{2}\Pi _i \left[ {1 - \frac{{m_\gamma ^2 }}
{{\nabla ^2 }}\left( {1 - \frac{{4m_H^2 }}{{\nabla ^2 }}} \right)}
\right]^{ - 1}\Pi ^ i} \right\}\left| \Phi  \right\rangle . \label{cond66}
\end{equation}
The reason why we eliminated from the effective Lagrangian the term involving $b_0$
now becomes clear, that is, the commutator for the fields $b_\mu$ is zero. Next,
it should be noted that expression (\ref{cond65}) may be conveniently rewritten as
\begin{equation}
\left\langle H \right\rangle _\Phi ^{\left( 1 \right)}  =  - \frac{1}{2}\frac{{4M^4 }}
{{\left( {M_2^2  - M_1^2 } \right)}}\int {d^3 } x\left\langle \Phi  \right|\Pi _i
\left\{ {\alpha \frac{{\nabla ^2 }}{{\left( {\nabla ^2  - M_1^2 } \right)}} -
\beta \frac{{\nabla ^2 }}{{\left( {\nabla ^2  - M_2^2 } \right)}}} \right\}
\Pi ^i \left| \Phi  \right\rangle, \label{cond70}
\end{equation}

with $\alpha  = \frac{1}{{\left( {M_1^2  - m_\gamma ^2 } \right)}}$ and $
\beta  = \frac{1}{{\left( {M_2^2  - m_\gamma ^2 } \right)}}$.  While $M_1^2  =
\frac{1}{2}\left( {m_\gamma ^2  + \sqrt {m_\gamma ^4  - 16M^4 } } \right)$,\\
$M_2^2  = \frac{1}{2}\left( {m_\gamma ^2  - \sqrt {m_\gamma ^4  - 16M^4 } } \right)$
and $M \equiv \sqrt {m_\gamma  m_H } $.

>From our above Hamiltonian analysis we observe that $\left\langle H \right\rangle _\Phi
^{\left( 1 \right)}$ takes the form
\begin{equation}
\left\langle H \right\rangle _\Phi ^{\left( 1 \right)}  = \left\langle H \right\rangle
_\Phi ^{\left( {1a} \right)}  + \left\langle H \right\rangle _\Phi ^{\left( {1b} \right)}, \label{cond75}
\end{equation}
where the $\left\langle H \right\rangle _\Phi ^{\left( {1a} \right)}$, $\left\langle H
\right\rangle _\Phi ^{\left( {1b} \right)}$ terms are given by
\begin{eqnarray}
\left\langle H \right\rangle _\Phi ^{\left( {1a} \right)}  &=&  - \frac{\alpha }{2}
\frac{{4M^4 }}{{\left( {M_2^2  - M_1^2 } \right)}}\int {d^3 x} \int_{\bf y}^
{{\bf y}^ \prime  } {dz_i^ \prime  } \delta ^{\left( 3 \right)} \left( {{\bf x} -
{\bf z}^ \prime  } \right)\left( {1 - \frac{{M_1^2 }}{{\nabla ^2 }}} \right)_x^{ - 1}
\times \nonumber \\&\times& \int_{\bf y}^{{\bf y}^ \prime  } {dz^i }
\delta ^{\left( 3 \right)} \left( {{\bf x} - {\bf z}} \right), \label{cond80a}
\end{eqnarray}
and
\begin{eqnarray}
\left\langle H \right\rangle _\Phi ^{\left( {1b} \right)}  &=&  \frac{\beta }{2}
\frac{{4M^4 }}{{\left( {M_2^2  - M_1^2 } \right)}}\int {d^3 x} \int_{\bf y}^
{{\bf y}^ \prime  } {dz_i^ \prime  } \delta ^{\left( 3 \right)} \left( {{\bf x} - {\bf z}^
\prime  } \right)\left( {1 - \frac{{M_2^2 }}{{\nabla ^2 }}} \right)_x^{ - 1}
\times \nonumber \\&\times& \int_{\bf y}^{{\bf y}^ \prime  } {dz^i }
\delta ^{\left( 3 \right)} \left( {{\bf x} - {\bf z}} \right). \label{cond80b}
\end{eqnarray}
One immediately sees that these expressions are analogous to that encountered in
previous works \cite{Gaete:2004dn,Gaete:2004ga,Gaete:2008qx,Gaete:2007zn,Gaete:2007fu}.
In view of this situation, we find that the potential for two opposite charges located
at ${\bf y}^\prime$ and $\bf y$ becomes
\begin{equation}
V =  - \frac{{q^2 }}{{4\pi }}\frac{{4M^4 }}{{\left( {M_2^2  - M_1^2 } \right)}}
\left[ {\frac{1}{{\left( {M_1^2  - m_\gamma ^2 } \right)}}\frac{{e^{ - M_1
|{\bf y} - {\bf y}^ \prime |} }}{{|{\bf y} - {\bf y}^ \prime  |}} - \frac{1}
{{\left( {M_2^2  - m_\gamma ^2 } \right)}}\frac{{e^{ - M_2 |{\bf y} - {\bf y}^
\prime  |} }}{{|{\bf y} - {\bf y}^ \prime |}} } \right].
\label{cond90}
\end{equation}
>From this expression we see that, in the limit $m_\gamma   \ll M$, the potential
reduces to
\begin{equation}
V = -\frac{{q^2 }}{{4\pi }}\frac{{e^{ - M|{\bf y} - {\bf y}^ \prime  |} }}
{{|{\bf y} - {\bf y}^ \prime  |}}\cos (M|{\bf y} - {\bf y}^ \prime  |). \label{cond95}
\end{equation}\\
Expression (\ref{cond95}) is identical to the one encountered in \cite{Gabadadze:2008pj},
which has been computed using  the propagator of the theory. Thus one is led to the
conclusion that the contributions of the gauge field propagator are properly captured
in the gauge invariant formalism.

\section{Higgs confining phase}

In this final part of the paper we consider a charged scalar field with a
``wrong-sign'' mass term and a quartic self-interaction potential. This is the simplest
model where the Higgs mechanism can occur. The new effect we are going to
study is the interplay between the Higgs vacuum and  the fermion condensate,
\begin{equation}
\mathcal{ L} =  - \frac{1}{4}F_{\mu \nu }^2  + \left| {D_\mu  \phi } \right|^2
+ m_H^2 \phi ^ *  \phi - \frac{\lambda}{6}\left(\, \phi ^ *  \phi\,\right)^2+e J^0
\delta^\mu_0\, A_\mu . \label{cond0555}
\end{equation}

Notice that the self-interaction coupling constant is assumed to be positive,
i.e. $\lambda>0$, in order to have a potential energy bounded from below. By setting
\begin{equation}
\phi\equiv \frac{\sigma}{\sqrt{2}}\, e^{i \alpha},
\end{equation}

and choosing the unitary gauge $\alpha=0$, we get

\begin{equation}
\mathcal{ L} =  - \frac{1}{4}F_{\mu \nu }^2+\frac{e^2 A^2 \sigma^2}{2}  +\frac{1}{2}
\left( \partial_\mu  \sigma  \right)^2  +\frac{m_H^2}{2}
\sigma^2 -\frac{\lambda}{24}\,\sigma^4+e J^0  \delta^\mu_0\, A_\mu + eJ^{\nu}_s A_{\nu}.
\label{cond055}
\end{equation}

The field equation obtained by varying (\ref{cond0555}) with respect to $A_\mu$ is
the same as before while the scalar field equation acquires a new term

\begin{eqnarray}
&& \partial_\mu F^{\mu\nu} +e^2 A^\nu \sigma^2 = eJ^\nu_s + e J^0
\delta^\nu_0, \label{f10}\\
&& \left(\, \Delta +e^2 A^2 + m^2_H - \frac{\lambda}{6}\, \sigma^2 \,\right)
\sigma =0. \label{f20}
\end{eqnarray}

We look for a homogeneous classical solution, as we did in the previous case

\begin{eqnarray}
&& \sigma=\mathrm{const.}\equiv \phi_0\ne 0\ , \label{cs10}\\
&& A_\mu \equiv \frac{\mu_s}{e}\,\delta^0_\mu. \label{cs20}
\end{eqnarray}

The two integration constants $\phi_0$, $\mu_s$ are determined by inserting
(\ref{cs10}), (\ref{cs20}) in (\ref{f10}), (\ref{f20}):
\begin{eqnarray}
&& e^2 A^\nu \phi^2_0 =   e J^0  \delta^\nu_0\label{f111}\\
&& \left(\, e^2\, A^2 + m^2_H -\frac{\lambda}{6} \phi^2_0 \,\right)
\phi_0 =0\label{f222}
\end{eqnarray}

Eq.(\ref{f111}) and (\ref{cs20}) determine $\mu_s$

\begin{equation}
\mu_s^3 + m^2_H \mu_s -\frac{\lambda}{6} J^0=0
\end{equation}

Eq.(\ref{cs10}) and (\ref{f222}) fix $\phi_0$

\begin{equation}
\phi_0^2=\frac{J^0}{\mu_s}
\end{equation}
To simplify calculation we consider the case $\mu_s<< m_H$. In this case we obtain

\begin{eqnarray}
&& \mu_s \approx \frac{\lambda J^0}{6 m_H^2}, \\
&& \phi_0^2 \approx  \frac{6 m_H^2}{\lambda }.
\end{eqnarray}

In summary, the  ground state of the system is described by the classical solution:

\begin{eqnarray}
&& \overline{\psi}_0 \gamma ^\mu \psi_0= \delta^\mu_0\, J^0\label{g11}, \\
&& \phi_0^2=\frac{J^0}{\mu_s}\approx \frac{6 m_H^2}{\lambda }\label{g22}, \\
&& A_{\mu }=\frac{\mu_s}{e} \,  \delta_\mu^0\approx\frac{\lambda J^0}{6 e m_H^2}\,
\delta_\mu^0 \label{g33}.
\end{eqnarray}

We write the fields as a Higgs vacuum expectation value and a quantum fluctuation:

\begin{eqnarray}
&& \sigma = \phi_0 +\eta\left(\, x\,\right)= \sqrt{ \frac{6 m_H^2}{\lambda } } +
\eta\left(\, x\,\right),  \label{fluc11}\\
&& A_\mu =\frac{\mu_s}{e} \,  \delta^0_\mu + b_\mu\left(\, x\,\right)=\frac{\lambda J^0}
{6 e m_H^2}\,  \delta_\mu^0+ b_\mu\left(\, x\,\right). \label{fluc22}
\end{eqnarray}

Next, we expand the Lagrangian up to quadratic terms in the fluctuations around the
Higgs vacuum

\begin{equation}
\mathcal{ L}^{(2)} =  - \frac{1}{4}f_{\mu \nu }^2
+ \frac{1}{2}m_\gamma ^2 b_\mu ^2  + 2e\phi_0 \mu_s\, b_0\,  \eta
+ \frac{1}{2}\left( {\partial _\mu  \eta } \right)^2 - m^2_H \eta^2
\ , \label{cond055b}
\end{equation}

where $m^2_\gamma\equiv e^2\phi_0^2\approx 6e^2 m^2_H/\lambda$.\\
As we anticipated in a previous section the scalar fluctuation  is now
\textit{massive} and integrating out the $\eta$ field induces a new effective
theory for the  $b_\mu$ field:

\begin{equation}
\mathcal{ L} =  - \frac{1}{4}f_{\mu \nu }^2  + \frac{1}{2}m_\gamma ^2 b_\mu ^2  +
\frac{1}{2}\, b_0
\frac{4e^2 \mu_s^2\, \phi_0^2}{\left( {\Delta  + 2m_H^2 } \right)}\, b_0 .
\label{cond100}
\end{equation}\\
In the same way as was done in the previous case, one finds
\begin{eqnarray}
\left\langle H \right\rangle _\Phi  &=&  - \frac{\alpha }{2}\int {d^3 x} \left\langle
\Phi  \right|\Pi _i \left\{ {\frac{1}{{M_1^2 }}\frac{{\nabla ^2 }}{{\left( {\nabla ^2
- M_2^2 } \right)}} - \frac{1}{{M_2^2 }}\frac{{\nabla ^2 }}{{\left( {\nabla ^2  - M_1^2 }
\right)}}} \right\}\Pi ^i \left| \Phi  \right\rangle  \nonumber \\
&+& \frac {\beta}{2} \int {d^3 x} \left\langle \Phi  \right|\Pi _i \left\{ {\frac{1}{{M_1^2 }}
\frac{1}{{\left( {\nabla ^2  - M_2^2 } \right)}} - \frac{1}{{M_2^2 }}\frac{1}
{{\left( {\nabla ^2  - M_1^2 } \right)}}} \right\}\Pi ^i \left| \Phi  \right\rangle,
\label{cond110}
\end{eqnarray}
where $\alpha  \equiv \frac{{6M^4 }}{{\left( {M_2^2  - M_1^2 } \right)}}$,
$\beta  \equiv \frac{{(2m_H^2) 6M^4}}{{\left( {M_2^2  - M_1^2 } \right)}}$.\\
While $M_1^2  = \frac{1}{2}\left[ {\left( {m_\gamma ^2  + 2m_H^2 } \right) +
 \sqrt {\left( {m_\gamma ^2  + 2m_H^2 } \right)^2  - 24M^4 } } \right]$, \\
$M_2^2  = \frac{1}{2}\left[ {\left( {m_\gamma ^2  + 2m_H^2 } \right) -
\sqrt {\left( {m_\gamma ^2  + 2m_H^2 } \right)^2  - 24M^4 } } \right]$ and
$M = \sqrt {m_\gamma  m_H }$.

Once again, following our earlier procedure \cite{Gaete:2004dn,Gaete:2004ga,Gaete:2008qx,Gaete:2007zn,Gaete:2007fu}, we
see that the potential for two opposite charges located at ${\bf y}^\prime$
and $\bf y$ takes the form
\begin{eqnarray}
V &=&  - \frac{{q^2 }}{{4\pi }}\alpha \left\{ {\frac{1}{{M_1^2 }}\frac{{e^{ - M_2
|{\bf y} - {\bf y}^ \prime  |} }}{{|{\bf y} - {\bf y}^ \prime  |}} - \frac{1}
{{M_2^2 }}\frac{{e^{ - M_1 |{\bf y} - {\bf y}^ \prime  |} }}{{|{\bf y} - {\bf y}^ \prime  |}}}
\right\} \nonumber \\
&+& \frac{{q^2 }}{{8\pi }}\beta \left\{ {\frac{1}{{M_1^2 }}\ln \left( {1 +
\frac{{\Lambda ^2 }}{{M_2^2 }}} \right) - \frac{1}{{M_2^2 }}\ln \left( {1 +
\frac{{\Lambda ^2 }}{{M_1^2 }}} \right)} \right\}|{\bf y} - {\bf y}^ \prime  |, \label{cond115}
\end{eqnarray}
where  $\Lambda$ is a short-distance cutoff. Here, in contrast to the previous case,
unexpected features are found. Interestingly, it is observed that the
$+ m_H^2 \phi ^ *  \phi$ term induces a Yukawa piece plus a linear
confining piece. Before going ahead, we would like to remark that the cutoff $\Lambda$ is a physical cutoff, so no ultraviolet divergence
arises in (\ref{cond115}). To understand why $\Lambda$ is a physical
cutoff, we observe that the first expression inside the second term on the right hand side of Eq.(\ref{cond115}) describes a flux tube, which is characterized by having a tension independent of its length. In fact, the electric field due to a charge leading to this flux is a constant electric field ($E_0$). In other words, the energy stored in the flux tube is proportional to its length. This implies that $\Lambda ^2  = M_2^2 \left( {e^{\frac{{8\pi M_1^2 }}{{q^2 \beta }}\left| {E_0 } \right|}  - 1} \right)$, which is finite. \\

\section{Final remarks}

In summary, we have considered the recently proposed Higgs-like model
\cite{Gabadadze:2008pj}, which describes a condensed of charged scalars
in a neutralizing background of fermions, from a somewhat different perspective.
First, we have studied the phenomenon of charged scalars in the familiar language
of standard quantum field theory. The internal consistency of this development was
illustrated. Second, a Hamiltonian analysis of the effective theory was done in order
to restore the gauge invariance. Third, based in the gauge-invariant but path-dependent
variables formalism, we have examined the confinement versus screening issue for this
new theory. When we compute in this way the static potential in the case of a
"right-sign" mass term $m_H^2 \phi ^ *  \phi$, we obtain an effective - Yukawa potential,
which in the  $m_\gamma   \ll M$ approximation is identical to the one encountered in \cite{Gabadadze:2008pj}. On the other hand, in the case of a "wrong-sign'' mass term
$- m_H^2 \phi ^ *  \phi$, the result of this calculation is new and rather unexpected.
We have showed that the interaction energy is the sum of an effective -Yukawa and a
\textit{linear potential}, leading to the confinement of static charges. As expressed
in the Introduction, similar forms of interaction potentials have been reported before
in different contexts. In this way a correspondence was established among diverse
effective models. The extension of these results to diverse dimensions or to
supersymmetric models would be welcome.

Acknowledgments\\

One of us (PG) wants to thank the Physics Department of the Universit\`a
di Trieste for hospitality and INFN for support. This work was
supported in part by Fondecyt (Chile) grant 1080260 (PG).

\end{document}